\documentclass[twocolumn]{autart}    
\usepackage{url}
\usepackage{xcolor}
\usepackage{amsmath,amssymb,amsfonts}
\usepackage{graphicx}          
\usepackage{cite}

\begin{document}

\begin{frontmatter}

\title{Composite Adaptive Disturbance Rejection in Robotics via Instrumental Variables based DREM\thanksref{footnoteinfo}} 

\thanks[footnoteinfo]{The material in this paper was not presented at any IFAC 
meeting. Corresponding author A.~Glushchenko. Tel. +79102266946.}

\author[Moscow]{Anton Glushchenko}\ead{aiglush@upi.ru},    
\author[Moscow]{Konstantin Lastochkin}\ead{lastconst@ipu.ru}               

\address[Moscow]{Laboratory 7, V. A. Trapeznikov Institute of Control Sciences of Russian Academy of Sciences, Moscow, Russia}  


\begin{abstract}                          
In this paper we consider trajectory tracking problem for robotic systems affected by unknown external perturbations. Considering possible solutions, we restrict our attention to composite adaptation, which, particularly, ensures parametric error convergence being desirable to enhance overall stability and robustness of a closed-loop system. At the same time, existing composite approaches cannot simultaneously relax stringent persistence of excitation requirement and guarantee convergence of parametric error to zero for a perturbed scenario. So, a new composite adaptation scheme is proposed, which successfully overcomes mentioned problems of known counterparts and has several salient features. First, it includes a novel adaptive disturbance rejection control law for a general \emph{n}-DoF dynamical model in the Euler-Lagrange form, which, without achievement of the parameter estimation goal, ensures global stability  via application of a high-gain external torque observer augmented with some adaptation law. Secondly, such law is extended with a composite summand derived via the recently proposed Instrumental Variables based Dynamic Regressor Extension and Mixing procedure, which relaxes excitation conditions and ensures asymptotic parameter estimation and reference tracking in the presence of external torque under some non-restrictive assumptions. An illustrative example shows the effectiveness and superiority of the proposed approach in comparison with existing solutions.
\end{abstract}

\end{frontmatter}

\section{Introduction}
Various types of robots are widely applied in industry to replace human labor to perform some dangerous, heavy or precise tasks \cite{b1}. Permanent strengthening of the accuracy specifications for the technological operations requires modern robotics to apply control laws, which have opportunity to handle uncertainties of environment and guarantee improved performance. In this regard, in recent years, one of the most popular theoretical developments, which is accompanied with relevant good sound practical examples, is the adaptive control with composite adaptation laws \cite{b2,b3,b4}.

In comparison with the well-known standard adaptive laws \cite{b5}, the composite adaptation is driven not only by the error of desired trajectory tracking, but also by extra model predictive terms obtained via some tedious filters-type transformations. This idea was initially proposed by K.S. Narendra \cite{b6} for model reference adaptive control, and independently by J.J. Slotine \cite{b7} for robotics. At the first stage of the historical development of composite adaptation, the extra terms allowed one to ({\it i}) speed up the convergence of both tracking and parameter estimation errors via inserting of high adaptation gains without excitation of high-frequency instabilities \cite{b7} and ({\it ii}) smooth the control transient response \cite{b8}. However, standard persistency of excitation (PE) condition was still required at this stage to achieve both convergence of parameter estimates to their ideal values and exponential stability of the overall closed-loop system. Exponential stability is a desirable property for adaptation and control laws as it leads to robustness (in the sense of input to state stability and uniform ultimate boundedness) against perturbations of various types, such as measurement noise, unmodeled dynamics, and external perturbations \cite{b9}. On the other hand, the parametric convergence in the adaptive systems can ensure accurate online modeling to solve high-level tasks, such as planning and diagnostics. So, at the second stage, the PE condition was relaxed, and the required exponential stability was ensured \cite{b10,b11,b12,b13,b14,b15}. All approaches for such relaxation are based on the intuitive idea to express unknown parameters from some linear regression equation (LRE) obtained on-line with the help of filters-type transformations and data-memory-driven or integral-like processing \cite{b2,b3,b4}. Typically, such composite adaptation laws are represented as the following differential equation:
\begin{displaymath}
\begin{array}{l}
\dot {\tilde \theta} \left( t \right) =  - \Gamma \left( {{\vartheta _b}\left( t \right) + {\vartheta _c}\left( t \right)} \right){\rm{,\;}}\hat \theta \left( {{t_0}} \right) = {{\hat \theta }_0}{\rm{,}}\\
{\vartheta _c}\left( t \right) = \gamma \left( {\Phi \left( t \right)\hat \theta \left( t \right) - Y\left( t \right)} \right){\rm{,}}
\end{array}    
\end{displaymath}
where $\tilde\theta\left( t \right)\!=\!\theta \!-\! \hat \theta\left( t \right)$ is the error of estimation of unknown parameters $\theta\in{\mathbb{R}^{{n_\theta }}}$, ${\vartheta _b}\left( t \right)\in{\mathbb{R}^{{n_\theta }}}$ is the baseline summand to compensate for the cross-terms in the Lyapunov analysis and achieve the trajectory tracking goal, ${\vartheta _c}\left( t \right) \in {\mathbb{R}^{{n_\theta }}}$ is a composite summand to ensure exponential stability and achieve parameters estimation, \linebreak $\Phi\left( t \right)\!=\!{\Phi^{\top}}\left( t \right)\in{\mathbb{R}^{{n_\theta} \times{n_\theta}}}$ denotes a regressor, $Y\left( t \right) \in {\mathbb{R}^{{n_\theta }}}$ stands for a regressand, $\Gamma  = {\Gamma ^{\top}} > 0{\rm{,\;}}\gamma  > 0$ are learning gains.

If the trajectory tracking goal is achieved, then the baseline summand ${\vartheta _b}\left( t \right)$ vanishes, and the estimation error is driven only by a composite term ${\vartheta _c}\left( t \right)$. Therefore, in the perturbation-free case, {\it i.e.} $Y\left( t \right) = \Phi \left( t \right)\theta $, it is obvious that the error $\tilde \theta \left( t \right)$ converges to zero exponentially if the regressor $\Phi \left( t \right)$ has a full rank. Aforementioned approaches \cite{b11,b12,b13,b14,b15} proposed other schemes to compute signals $Y\left( t \right)$ and $\Phi \left( t \right)$ on the basis of the generalized position and velocity of a robot such that the regressor $\Phi \left( t \right)$ has full rank without restrictive PE requirement. Interested readers are referred to recent reviews \cite{b2,b3,b4} to be provided with a deep insight into the schemes under consideration.

In a more complicated perturbed scenario, {\it i.e.} \linebreak $Y\left( t \right)\!=\!\Phi \left( t \right)\theta \!+\! W\left( t \right)$, considering the existing solutions, the uniform ultimate boundedness (UUB) of both tracking and parametric errors is usually proved (see Theorems 5.1, 7.1 in \cite{b10} and Theorem 2 in \cite{b12}). Unfortunately, such results are fully impracticable as:
\begin{enumerate}
    \item[{\it i})] the ultimate bounds can be arbitrary large with no opportunity to adjust them via the learning gains and other parameters of both control and adaptation laws,
    \item[{\it ii})] in reality, perturbation always exists and affects the system, {\it e.g.} an external torque for a robot.
\end{enumerate}

As a simple example, it is assumed that the approach from \cite{b12} is applied to obtain static regressor and perturbation such that:
\begin{displaymath}
\forall t \ge {t_0}{\rm{\;}}\Phi \left( t \right): = {\begin{bmatrix}
{{\Phi _{11}}}&0\\
0&{{\Phi _{12}}}
\end{bmatrix}} {\rm{,\;}}W\left( t \right){\rm{:}} = {\begin{bmatrix}
{0.1}\\
{ - 0.1}
\end{bmatrix}}.    
\end{displaymath}
Then it is easy to see that the equilibrium point of the parametric error $\tilde \theta \left( t \right)$ is
\begin{displaymath}
{\tilde \theta _{eq}}{\rm{:}} = {\begin{bmatrix}
{\Phi _{11}^{ - 1}}&0\\
0&{\Phi _{22}^{ - 1}}
\end{bmatrix}}{\begin{bmatrix}
{0.1}\\
{ - 0.1}
\end{bmatrix}} = {\begin{bmatrix}
{0.1\Phi _{11}^{ - 1}}\\
{ - 0.1\Phi _{22}^{ - 1}}
\end{bmatrix}}{\rm{,}}    
\end{displaymath}
which means that in the presence of external perturbation the uniform ultimate bound for $\tilde \theta \left( t \right)$ can probably be arbitrarily large when $\Phi \left( t \right)$ is ill-conditioned.

In practice, the disturbance always comes along with the motion process of robotic systems. Therefore, it is necessary to take the external disturbance into account in case the trajectory tracking control problem is considered. However, to the best of authors’ knowledge, in the existing literature on composite adaptive control, there is no composite summand ${\vartheta _c}\left( t \right)$, which simultaneously relaxes PE condition and guarantees convergence of $\tilde \theta \left( t \right)$ to zero in case of the perturbed scenario. Therefore, our contribution is that a composite adaptation scheme to overcome this drawback is developed for the first time.

To achieve the stated goal, we decompose the control design into two parts. Firstly, the novel adaptive disturbance rejection control law for a general $n$-DoF dynamical model in the Euler-Lagrange form is proposed, which ensures disturbance rejection and adaptive regulation. Global stability is achieved via a high-gain external torque observer augmented with the baseline adaptation law to compensate for some linearly parameterized cross terms in the Lyapunov analysis. This implies that the acceptable stable behavior of the robot is guaranteed without achievement of the parameter estimation goal. At the second stage, the recently proposed instrumental variables based Dynamic Regressor Extension and Mixing procedure (IV based DREM \cite{b16}) is adopted for robotics to obtain composite summand, which relaxes excitation conditions and ensures asymptotic parameter estimation and reference tracking in the presence of external torque. In its turn, the instrumental variable is chosen as a system regressor calculated via substitution of the reference trajectory signals. Then the extension and mixing procedure based on the instrumental variable, sliding window and averaging filters is applied to obtain a set of independent scalar linear regression equations with asymptotically vanishing and square integrable perturbation. Afterwards, based on such a set of equations, a composite summand is calculated via gradient descent method. If ({\it i}) the weighted rate of the external torque is square integrable, ({\it ii}) the external torque affecting the plant is independent with the system trajectory ({\it e.g.} in a stationary case their spectra do not contain common frequencies), ({\it iii}) the scalar regressor is not integrable with square, then the proposed composite summand ensures asymptotic convergence of the parameter estimation and tracking errors to zero. Moreover, it is shown that the requirement of non-square integrability is weaker than the PE one. \emph{The significance and distinction of this study is that the parameter estimation objective in composite adaptive control scheme is achieved with relaxed excitation requirements and for robots affected by external unknown torque.}

In the remainder of this manuscript, Section 2 formulates the control problem, Section 3 presents the proposed control design, Section 4 provides illustrative results, and Section 5 draws conclusions.

\textbf{Notation.} Further the following notation is used: \linebreak $\mathbb{R}_{>0}$
denote the positive real numbers, $\left| . \right|$ is the absolute value, $\left\| . \right\|$ is the suitable norm of $(.)$, ${I_{n \times n}}=I_{n}$ is an identity $n \times n$ matrix, ${0_{n \times n}}$ is a zero $n \times n$ matrix, \linebreak $0_{n}$ stands for a zero vector of length $n$, ${\rm{det}}\{.\}$ stands for a determinant, ${\rm{adj}}\{.\}$ represents an adjoint. {We say that $f\in L_{q}$ if  $\sqrt[q]{{\int\limits_{{t_0}}^t {{\left|f\left( s \right)\right|^q}ds} }} < \infty $ for all $t\ge t_{0}$.}

\section{Problem Statement}\label{s2}
Consider a standard $n$-DoF dynamical model in the Euler-Lagrange form\footnote{Hereafter, whenever it causes no confusion, the argument $t$ will be omitted.}:
\begin{equation}\label{eq1}
\begin{gathered}
M\left( q \right)\ddot q + C\left( {q{\rm{,\;}}\dot q} \right)\dot q + F\left( {\dot q} \right) + G\left( q \right) = \tau + {\tau _d}{\rm{,}}    
\end{gathered}
\end{equation}
where $q{\rm{,\;}}\dot q{\rm{,\;}}\ddot q \in {\mathbb{R}^n}$ denote generalized position, velocity and acceleration, respectively, $M{\rm{:\;}}{\mathbb{R}^n} \mapsto {\mathbb{R}^{n \times n}}$ is a generalized inertia matrix, $C{\rm{:\;}}{\mathbb{R}^n} \times {\mathbb{R}^n}  \mapsto {\mathbb{R}^{n \times n}}$ stands for a Coriolis and centrifugal forces matrix, $F{\rm{:\;}}{\mathbb{R}^n} \mapsto {\mathbb{R}^n}$ and $G{\rm{:\;}}{\mathbb{R}^n} \mapsto {\mathbb{R}^n}$ represent friction and gravitation effects, $\tau \in {\mathbb{R}^n}$ is a control torque and ${\tau _d} \in {\mathbb{R}^n}$ is an external unknown but bounded torque $\left\| {{\tau _d}\left( t \right)} \right\| \le \mathop {{\rm{sup}}}\limits_t \left\| {{\tau _d}\left( t \right)} \right\| < \infty $. It is assumed that the states $q{\rm{,\;}}\dot q$ are measurable and the system \eqref{eq1} belongs to the class defined by the following properties.

\textbf{Property 1.} For any auxiliary vector $v \in {\mathbb{R}^n}$ the left side of the system \eqref{eq1} admits linear parameterizations
\begin{displaymath}
\begin{array}{l}
M\left( q \right)\dot v + C\left( {q{\rm{,\;}}\dot q} \right)v + F\left( {\dot q} \right)
+G\left( q \right){\rm{:}} = {\Phi ^{\top}}\left( {q{\rm{,\;}}\dot q{\rm{,\;}}v{\rm{,\;}}\dot v} \right)\theta {\rm{,}}\\
{\Phi ^{\top}}\left( {q{\rm{,\;}}\dot q{\rm{,\;}}v{\rm{,\;}}\dot v} \right){\rm{:}} = \Phi _M^{\top}\left( {q{\rm{,\;}}\dot v} \right) + \Phi _C^{\top}\left( {q{\rm{,\;}}\dot q{\rm{,\;}}v} \right) +\\\hfill
+\Phi _F^{\top}\left( {\dot q} \right) + \Phi _G^{\top}\left( q \right){\rm{,}}\\
\Phi _{\partial M}^{\top}\left( {q{\rm{,\;}}\dot q{\rm{,\;}}\dot v} \right)\theta {\rm{:}} = \dot M\left( q \right)\dot v{\rm{,}}
\end{array}    
\end{displaymath}
where ${\Phi ^{\top}}\left( {q{\rm{,\;}}\dot q{\rm{,\;}}v{\rm{,\;}}\dot v} \right) \in {\mathbb{R}^{n \times {n_\theta }}}$ is a regressor, $\theta  \in {\mathbb{R}^{{n_\theta }}}$ is a vector of unknown time-invariant parameters.

\textbf{Property 2.} The inertia matrix $M\left( q \right)$ is symmetric, positive-definite and satisfies the following inequality
\begin{displaymath}
\underline m {I_n} \le M\left( q \right) \le \overline m{I_n}{\rm{,}}    
\end{displaymath}
for some $0 < \underline m  \le \overline m$.

\textbf{Property 3.} The inertia and Coriolis matrices meet the skew-symmetric relation:
\begin{displaymath}
{x^{\top}}\left( {{\textstyle{1 \over 2}}{\textstyle{d \over {dt}}}\left[ {M\left( q \right)} \right] - C\left( {q{\rm{,\;}}\dot q} \right)} \right)x = 0{\rm{,\;}}\forall x \in {\mathbb{R}^n}.    
\end{displaymath}
\textbf{Property 4.} There exists a known continuous differentiable function $\mu {\rm{:\;}}\left[ {{t_0}{\rm{,\;}}\infty } \right) \mapsto {\mathbb{R}_{>0} }$ such that:
\begin{enumerate}
    \item[\textbf{4.1)}] the weighted rate $\lambda = {\mu ^{ - 1}}{\dot \tau _d}$ is square integrable, \emph{i.e.} $\lambda  \in {L_2}$:
    \begin{equation*}
    \begin{array}{l}\hspace{-25pt}
\sqrt {\int\limits_{{t_0}}^t {{{\left\| {\lambda \left( s \right)} \right\|}^2}ds} }  = \sqrt {\int\limits_{{t_0}}^t {{\mu ^{ - 2}}\left( s \right){{\left\| {{{\dot \tau }_d}\left( s \right)} \right\|}^2}ds} }  < \infty {\rm{,\;}}\forall {{t}}  \ge {t_0}{\rm{,}}        
    \end{array}
       \end{equation*}
    \item[\textbf{4.2)}] for a known constant ${\delta _\mu } \in \left( {0,{\rm{ 1}}} \right)$ there exists a function ${\mu _{{\rm{LB}}}}{\rm{:\;}}\left[ {{t_0}{\rm{,\;}}\infty } \right) \mapsto {\mathbb{R}_{>0} }$ and values \linebreak $\rho  > 0{\rm{,\;}}{\delta _\rho } > 0$ such that
    \begin{displaymath}
\left( {\rho  - \delta _\mu ^{ - 1}} \right)\mu \left( t \right) + \rho {\textstyle{{\dot \mu \left( t \right)} \over {\mu \left( t \right)}}} - \rho \delta _\rho ^{ - 1} \ge {\mu _{{\rm{LB}}}}\left( t \right) > 0{\rm{,\;}}\forall {{t}}  \ge {t_0}, 
    \end{displaymath}
    and ${\mu _{{\rm{LB}}}} \notin {L_1}$.
\end{enumerate}

The goal is to track the continuous and twice differentiable trajectory ${q_d} \in {\mathbb{R}^n}$ such that the following equalities hold:
\begin{equation}\label{eq2}
\mathop {{\rm{lim}}}\limits_{t \to \infty } \left\| {e\left( t \right)} \right\| = 0{\rm{,\;}}\mathop {{\rm{lim}}}\limits_{t \to \infty } \left\| {\tilde \theta \left( t \right)} \right\| = 0{\rm{,}}    
\end{equation}
where $e\left( t \right) = {q_d}\left( t \right) - q\left( t \right)$, $\tilde \theta \left( t \right) = \theta  - \hat \theta \left( t \right)$ are tracking and parametric errors, respectively.

Moreover, the convergence of $\tilde \theta \left( t \right)$ to zero needs to be achieved simultaneously:
\begin{enumerate}
    \item[{\it i})] in the presence of external unknown torque,
    \item[{\it ii})] without restrictive persistence of excitation assumption.
\end{enumerate}

\textbf{Remark 1.} Considering practical scenarios, it almost always holds that ${\dot \tau _d} \in {L_2}$, so in most cases it is sufficient to select $\mu \left( t \right) = \mu  > 0$. In case of perturbation with ${\dot \tau _d} \notin {L_2}$, but $\left\| {{{\dot \tau }_d}} \right\| \le {c_{{\tau _d}}} < \infty $, according to the mean value theorem, the following inequality holds for any monotonous weighted function
\begin{equation*}
\begin{array}{l}
\sqrt {\int\limits_{{t_0}}^t {{\mu ^{ - 2}}\left( s \right){{\left\| {{{\dot \tau }_d}\left( s \right)} \right\|}^2}ds} }  \le {c_{{\tau _d}}}\sqrt {\int\limits_{{t_0}}^t {{\mu ^{ - 2}}\left( s \right)ds} } {\rm{,\;}}\forall {{t}}  \ge {t_0}{\rm{,}}    
\end{array}
\end{equation*}
and, therefore, the straightforward choice is $\mu \left( t \right) = {\mu _0}t +\\+ {\mu _1}$ for some ${\mu _0} > 0{\rm{,\;}}{\mu _1} > 0$. For $\mu \left( t \right) = \mu  > 0$ and $\mu \left( t \right) = {\mu _0}t + {\mu _1}$ , it is obvious that there exist $\rho  \!>\! 0{\rm{,\;}}{\delta _\rho } \!>\! 0$ such that property 4.2 is satisfied for all ${\delta _\mu } \in \left( {0,{\rm{ 1}}} \right)$.

Unfortunately, for perturbations ${\tau _d}$ with unbounded change rate, some extra {\it a priori} knowledge is required to choose an appropriate weighted function $\mu \left( t \right)$.

\section{Main Result}\label{s3}

The main result of the study is presented sequentially in three sections. In Section 3.1, a novel control law with composite adaptive disturbance rejection scheme is proposed. In Section 3.2, a new composite summand for the base adaptation law is derived on the basis of the method from \cite{b16}. Section 3.3 is dedicated to the stability analysis of the closed-loop control system.

\subsection{Control Law Design}

To design the control law, a filtered error $r \!=\! \dot e + \alpha e{\rm{,\;}}\alpha  \!>\! {\rm{0}}$ is differentiated with respect to time, and the obtained result is multiplied by $M\left( q \right)$:
\begin{equation}\label{eq3}
\begin{array}{l}
M\left( q \right)\dot r = M\left( q \right)\left[ {\ddot e + \alpha \dot e} \right] = \hfill\\
 = M\left( q \right)\left[ {{{\ddot q}_d} + \alpha \dot e} \right] - M\left( q \right)\ddot q = \hfill\\
 = M\left( q \right)\dot v + C\left( {q{\rm{,\;}}\dot q} \right)\dot q + F\left( {\dot q} \right)
 +G\left( q \right) - \tau - {\tau _d}{\rm{,}}
\end{array}    
\end{equation}
where $v = {\dot q_d} + \alpha e$ and, consequently, $\dot v  = {\ddot q_d} + \alpha \dot e$.

Substitution $\dot q = v - r$ and application of property 1 result in the following equation:
\begin{equation}\label{eq4}
\begin{gathered}
M\left( q \right)\dot r = {\Phi ^{\top}}\left( {q{\rm{,\;}}\dot q{\rm{,\;}}v{\rm{,\;}}\dot v} \right)\theta 
- C\left( {q{\rm{,\;}}\dot q} \right)r - \tau - {\tau _d}. 
\end{gathered}
\end{equation}
It is obvious from equation \eqref{eq4} that, to achieve the tracking objectives \eqref{eq2}, both parametric uncertainty and external torque terms need to be handled. To design a control law, firstly, we derive a high gain external torque observer. To do that the following auxiliary state is introduced:
\begin{equation}\label{eq5}
x : = M\left( q \right)r.    
\end{equation}
Then, owing to property 1, equation \eqref{eq4} could be rewritten in a simpler form:
\begin{equation}\label{eq6}
\begin{array}{l}
\dot x\left( t \right) = \left( {\Phi ^{\top}}\left( {q{\rm{,\;}}\dot q{\rm{,\;}}v{\rm{,\;}}\dot v} \right) - \Phi _C^{\top}\left( {q{\rm{,\;}}\dot q{\rm{,\;}}r} \right) + \right. \\\quad
\left. + \Phi _{\partial M}^{\top}\left( {q{\rm{,\;}}\dot q{\rm{,\;}}r} \right) \right)\theta  + f\left(t\right) + u\left(t\right){\rm{,\;}}x\left( {{t_0}} \right) = {x_0}{\rm{,}}
\end{array}    
\end{equation}
where the following notation is used:
\begin{displaymath}
u\left( t \right){\rm{:}} =  - \tau \left( t \right){\rm{,\;}}\;f\left( t \right){\rm{:}} =  - {\tau _d}\left( t \right).    
\end{displaymath}
Temporarily suppose that ${\dot\tau _d}\in{L_2}$, then $\mu \left( t \right)\!=\!\mu \! >\!0$ (Remark 1) and the filtered version of unknown input $f\left( t \right)$ meets the following equation:
\begin{equation*}
\begin{array}{l}
\hat f\left( t \right){\rm{:}} = \frac{1}{{{\mu ^{ - 1}}s + 1}}\left[ {f}\right]\left( t \right) =\\=  \frac{s}{{{\mu ^{ - 1}}s + 1}}\left[ {x} \right] \left( t \right)
- \frac{1}{{{\mu ^{ - 1}}s + 1}}\left[ {u} \right]\left( t \right) -\hfill \\ \hfill- \frac{1}{{{\mu ^{ - 1}}s + 1}}\left[ {\Phi ^{\top}} - \Phi _C^{\top} + \Phi _{\partial M}^{\top} \right]\left( t \right)\theta {\rm{,}}
\end{array}    
\end{equation*}
where ${\textstyle{1 \over {{\mu ^{ - 1}}s + 1}}}\left[ . \right]\left( t \right)$ is a stable low pass filter ($s{\rm{:}} = {\textstyle{d \over {dt}}}$).

The above-obtained filtered perturbation for ${\dot \tau _d} \in {L_2}$ motivates to consider a more general case ${\dot \tau _d} \notin {L_2}$ and introduce the following high-gain unknown input observer:
\begin{equation}\label{eq7}
\begin{array}{l}
\hat f\left( t \right){\rm{:}} = \chi \left( t \right) - \Phi _f^{\top}\left( t \right)\theta  - {u_f}\left( t \right) = \\
 =\! \mu\! \left( t \right)\!\left[ {\Phi _M^{\top}\left( {q{\rm{,\;}}r} \right) \!-\! \Phi _{Mf}^{\top}\left( t \right) \!-\! {\textstyle{1 \over \mu\left(t\right) }}\Phi _f^{\top}\left( t \right)} \right]\!\theta  \!-\! {u_f}\!\left( t \right){\rm{,}}
\end{array}
\end{equation}
where
\begin{displaymath}
\begin{array}{l}
\chi \left( t \right) = \mu \left( t \right)\left( {x\left( t \right) - {x_f}\left( t \right)} \right){\rm{,}}\\
{{\dot x}_f}\left( t \right) = \left( {\mu \left( t \right) + {\textstyle{{\dot \mu \left( t \right)} \over {\mu \left( t \right)}}}} \right)\left( {x\left( t \right) - {x_f}\left( t \right)} \right){\rm{,\;}}{x_f}\left( {{t_0}} \right) = {0_n}{\rm{,}}\\
{{\dot \Phi }_f}\left( t \right) = \mu \left( t \right)\left( {\Phi ^{\top}}\left( {q{\rm{,\;}}\dot q{\rm{,\;}}v{\rm{,\;}}\dot v} \right) - \Phi _C^{\top}\left( {q{\rm{,\;}}\dot q{\rm{,\;}}r} \right) + \right.\\
\left. + \Phi _{\partial M}^{\top}\left( {q{\rm{,\;}}\dot q{\rm{,\;}}r} \right) - {\Phi _f}\left( t \right) \right){\rm{,\;}}{\Phi _f}\left( {{t_0}} \right) = {0_{{n_\theta } \times n}}{\rm{,}}\\
{{\dot \Phi }_{Mf}}\!\left( t \right) \!=\! \mu \left( t \right)\left( {{\Phi _M}\left( {q{\rm{,\;}}r} \right) \!-\! {\Phi _{Mf}}\left( t \right)} \right){\rm{,\;}}{\Phi _{Mf}}\left( {{t_0}} \right) \!=\! {0_{{n_\theta } \times n}}{\rm{,}}\\
{{\dot u}_f}\left( t \right) = \mu \left( t \right)\left( {u\left( t \right) - {u_f}\left( t \right)} \right){\rm{,\;}}{u_f}\left( {{t_0}} \right) = {0_n}.
\end{array}    
\end{displaymath}
Therefore, substitution of \eqref{eq7} into \eqref{eq4} results in the following dynamics:
\begin{equation}\label{eq8}
\begin{array}{l}
M\left( q \right)\dot r = \\ = {\Phi ^{\top}}\left( {q{\rm{,\;}}\dot q{\rm{,\;}}v{\rm{,\;}}\dot v} \right)\theta  - C\left( {q{\rm{,\;}}\dot q} \right)r
- \tau - {\tau _d} \pm \hat f = \\ =  - \tau - C\left( {q{\rm{,\;}}\dot q} \right)r
+ {\Pi ^{\top}}\theta  + \mu \left( t \right){{\tilde f}_n} - {u_f}{\rm{,}}
\end{array}    
\end{equation}
where ${\tilde f_n}\left( t \right) = {\textstyle{1 \over {\mu \left( t \right)}}}\left( {f\left( t \right) - \hat f\left( t \right)} \right)$ is normalized unknown input reconstruction error and ${\Pi ^{\top}}\left( t \right) \in {\mathbb{R}^{n \times {n_\theta }}}$ is a new regressor:
\begin{equation*}
\begin{array}{l}
{\Pi ^{\top}}\left( t \right){\rm{:}} = \mu \left( t \right)\left( {\Phi _M^{\top}\left( {q{\rm{,\;}}r} \right) - \Phi _{Mf}^{\top}\left( t \right)} \right) + \hfill\quad\quad\quad\quad\quad\quad\\\hfill
+ {\Phi ^{\top}}\left( {q{\rm{,\;}}\dot q{\rm{,\;}}v{\rm{,\;}}\dot v} \right) - \Phi _f^{\top}\left( t \right).    
\end{array}
\end{equation*}
Now, we are in position to define the control law as follows: 
\begin{equation}\label{eq9}
\tau = \left( {K + {\delta _\mu }\mu \left( t \right)} \right)r + {\Pi ^{\top}}\hat \theta - {u_f}{\rm{,}}    
\end{equation}
where $K \in {\mathbb{R}^{n \times n}}$ is a positive definite diagonal matrix, and ${\delta _\mu } \in \left( {0,{\rm{ 1}}} \right)$ is a value from property 4.

The second and third summands of \eqref{eq9} are designed to compensate for both external torque and parametric uncertainty in the closed loop equation \eqref{eq4}.

After substitution of the control torque \eqref{eq9} into \eqref{eq8}, the closed-loop filtered tracking error dynamics yields:
\begin{equation}\label{eq10}
\begin{array}{l}
M\left( q \right)\dot r = \\\; - \left( {K + {\delta _\mu }\mu \left( t \right)} \right)r
- C\left( {q{\rm{,\;}}\dot q} \right)r + {\Pi ^{\top}}\tilde \theta  + \mu \left( t \right){\tilde f_n}.
\end{array}
\end{equation}
Based on the closed-loop equation \eqref{eq10}, the general composite adaptation law is introduced:
\begin{equation}\label{eq11}
\dot {\hat \theta} \left( t \right) = \Gamma \left( {{\Pi ^{\top}}\left( t \right)r\left( t \right) + {\vartheta _c}\left( t \right)} \right){\rm{,\;}}\hat \theta \left( {{t_0}} \right) = {\hat \theta _0}{\rm{,}}    
\end{equation}
where $\Gamma  = {\Gamma ^{\top}} > 0$ is a learning rate, and ${\vartheta _c}\in {\mathbb{R}^{{n_\theta }}}$ is an extra term to be applied in the adaptation law based on the other sources of parametric information.

For example, if we select ${\vartheta _c}\left( t \right) = \sigma \left( {\hat \theta \left( t \right) - {\theta ^ * }} \right)$, where $\sigma  > 0{\rm{,}}\;\left\| {{\theta ^ * }} \right\| > 0$, then it is easy to prove that the errors $e\left( t \right)$ and $\tilde \theta \left( t \right)$ are uniformly ultimately bounded in some small residual set, which size could not be reduced arbitrarily \cite{b5}. Some schemes to select ${\vartheta _c}\left( t \right)$ based on the dynamic memory, which is designed via regressor extension and integral-like operations, could be found in the recent exhaustive reviews \cite{b2,b3,b4}. In this study, the Instrumental Variables based DREM procedure \cite{b16} is adopted to propose novel approach to choose ${\vartheta _c}\left( t \right)$ such that the stated goal \eqref{eq2} is achieved. In comparison with the existing approaches, the contribution of the latter is to guarantee convergence of $\tilde \theta \left( t \right)$ to zero: {\it i}) in the presence of non-vanishing external torque, and {\it ii}) without restrictive persistence of excitation assumption.

\textbf{Remark 2}. If all used filters are low-pass, then it can be clearly seen from the obtained closed-loop equation \eqref{eq10} and regressor $\Pi \left( t \right)$ definition that the total closed-loop uncertainty exists only at relatively high frequencies. Therefore, the high values of learning rate $\Gamma $ is not necessary to achieve a relatively good tracking response in the proposed composite adaptive disturbance rejection control scheme.

\subsection{Novel Scheme of Composite Adaptation}

The model \eqref{eq1} that satisfies property 1 could be rewritten in the following linear regression equation form:
\begin{equation}\label{eq12}
\tau \left( t \right) = {\Phi ^{\top}}\left( {q{\rm{,\;}}\dot q{\rm{,\;}}\ddot q} \right)\theta  - {\tau _d}\left( t \right).    
\end{equation}
Then the filtered version of \eqref{eq12} could be obtained
\begin{equation}\label{eq13}
z\left( t \right) = {\varphi ^{\top}}\left( t \right)\theta  + w\left( t \right){\rm{,}}
\end{equation}
where ($l > 0$)
\begin{displaymath}
\begin{array}{l}
\dot z\left( t \right) = l\left( {\tau \left( t \right) - z\left( t \right)} \right){\rm{,\;}}z\left( {{t_0}} \right) = {0_n}{\rm{,}}\\
\dot \varphi \left( t \right) = l\left( {\Phi \left( {q{\rm{,\;}}\dot q{\rm{,\;}}\ddot q} \right) - \varphi \left( t \right)} \right){\rm{,\;}}\varphi \left( {{t_0}} \right) = {0_{{n_\theta } \times n}}
\end{array}    
\end{displaymath}
and the filtered external torque satisfies
\begin{displaymath}
\dot w\left( t \right) = l\left( { - {\tau _d}\left( t \right) - w\left( t \right)} \right){\rm{,\;}}w\left( {{t_0}} \right) = {0_n}.    
\end{displaymath}

It should be noted that the regressor $\varphi \left( t \right)$ could be computed without knowledge of $\ddot q$ via integration by parts rule (for more details see \cite{b7} or Chapter 6, p. 362 of \cite{b1}).

Following the method from \cite{b16}, an instrumental variable is introduced:
\begin{equation}\label{e14}
\dot \zeta \left( t \right) = l\left( {\Phi \left( {{q_d}{\rm{,\;}}{{\dot q}_d}{\rm{,\;}}{{\ddot q}_d}} \right) - \zeta \left( t \right)} \right){\rm{,\;}}\zeta \left( {{t_0}} \right) = {0_{{n_\theta } \times n}}{\rm{,}}    
\end{equation}
and equation \eqref{eq13} is transformed as follows:
\renewcommand{\theequation}{\arabic{equation}a}
\begin{equation}\label{eq15a}
\begin{array}{l}
\dot y\left( t \right) = \zeta \left( t \right)z\left( t \right) - \zeta \left( {t - T} \right)z\left( {t - T} \right){\rm{, }}\\
\dot \psi \left( t \right) = \zeta \left( t \right){\varphi ^{\top}}\left( t \right) - \zeta \left( {t - T} \right){\varphi ^{\top}}\left( {t - T} \right){\rm{, }}\\
y\left( {{t_0}} \right) = {0_{{n_\theta }}}{\rm{,\;}}\psi \left( {{t_0}} \right) = {0_{{n_\theta } \times {n_\theta }}}{\rm{,}}
\end{array}
\end{equation}
\setcounter{equation}{14}
\renewcommand{\theequation}{\arabic{equation}b}
\begin{equation}\label{eq15b}
\begin{array}{l}
\dot Y\left( t \right) =  - \frac{1}{{F\left( t \right)}}\dot F\left( t \right)\left( {Y\left( t \right) - y\left( t \right)} \right){\rm{, }}\\
\dot \Psi \left( t \right) =  - \frac{1}{{F\left( t \right)}}\dot F\left( t \right)\left( {\Psi \left( t \right) - \psi \left( t \right)} \right){\rm{, }}\\
\dot F\left( t \right) = p{t^{p - 1}}{\rm{, }}\\
Y\left( {{t_0}} \right) = {0_{{n_\theta }}}{\rm{,\;}}\Phi \left( {{t_0}} \right) = {0_{{n_\theta } \times {n_\theta }}}{\rm{,\;}}F\left( {{t_0}} \right) = {F_0}{\rm{,}}
\end{array}
\end{equation}
where $T > 0$ is a sliding window width, $p \ge 1$ and ${F_0} \ge t_0^p$ stand for the filter parameters.

According to proposition 4 from \cite{b16}, the following regression equation is obtained as a result of the transformations \eqref{eq15a}, \eqref{eq15b}:
\renewcommand{\theequation}{\arabic{equation}}
\setcounter{equation}{15}
\begin{equation}\label{eq16}
Y\left( t \right) = \Psi \left( t \right)\theta  + W\left( t \right){\rm{,}}    
\end{equation}
where the new disturbance $W\left( t \right)$ satisfies the equation:
\begin{displaymath}
\begin{array}{l}
\dot W\left( t \right) =  - \frac{1}{{F\left( t \right)}}\dot F\left( t \right)\left( {W\left( t \right) - \varepsilon \left( t \right)} \right){\rm{,\;}}W\left( {{t_0}} \right) = {0_{{n_\theta }}}{\rm{,}}\\
\dot \varepsilon \left( t \right) = \zeta \left( t \right)w\left( t \right) - \zeta \left( {t - T} \right)w\left( {t - T} \right){\rm{,\;}}\varepsilon \left( {{t_0}} \right) = {0_{{n_\theta }}}.
\end{array}    
\end{displaymath}
Following the dynamic regressor extension and mixing procedure, the multiplication of \eqref{eq16} by ${\rm{adj}}\left\{ {{\textstyle{{\Psi \left( t \right)} \over {1 + \left\| {\Psi \left( t \right)} \right\|}}}} \right\}$ yields:
\begin{equation}\label{eq17}
{{\mathcal Y}_i}\left( t \right) = \Delta \left( t \right){\theta _i} + {{\mathcal W}_i}\left( t \right){\rm{,}}    
\end{equation}
where
\begin{displaymath}
\begin{array}{c}
{\mathcal Y}\left( t \right){\rm{:}} = {\rm{adj}}\left\{ {{\textstyle{{\Psi \left( t \right)} \over {1 + \left\| {\Psi \left( t \right)} \right\|}}}} \right\}Y\left( t \right){\rm{, }}\\
\Delta \left( t \right)\!{\rm{:}} \!=\! {\rm{det}}\left\{ {{\textstyle{{\Psi \left( t \right)} \over {1 + \left\| {\Psi \left( t \right)} \right\|}}}} \right\}{\rm{,\;}}{\mathcal W}\left( t \right)\!{\rm{:}} \!=\! {\rm{adj}}\left\{ {{\textstyle{{\Psi \left( t \right)} \over {1 + \left\| {\Psi \left( t \right)} \right\|}}}} \right\}W\left( t \right)
\end{array}    
\end{displaymath}
and $i = {1, \ldots {\rm{,}}{n_\theta }}$.

The normalization operation is used to guarantee the boundedness of all signals in equation \eqref{eq17} even in case of unbounded tracking error $e\left( t \right)$. Moreover, the following propositions hold for the regressor $\Delta \left( t \right)$ and disturbance ${\mathcal W}\left( t \right)$.

\textbf{Proposition 1.} \emph{If for the picked sliding window $T > 0$ there exists $\beta  > 0$ such that}
\begin{equation}\label{eq18}
\left| {{\rm{det}}\left\{ {\int\limits_t^{t + T} {\zeta \left( \tau  \right){\varphi ^{\top}}\left( \tau  \right)d\tau } } \right\}} \right| \ge \beta  > 0{\rm{,\;}}\forall t \ge {t_0}{\rm{,}}    
\end{equation}
\emph{then also there exists a time instant ${T_\Delta } \ge {t_0} + T$ and value ${\Delta _{{\rm{LB}}}} > 0$ such that $\left| {\Delta \left( t \right)} \right| \ge {\Delta _{{\rm{LB}}}} > 0$ for all $t \ge {T_\Delta }$.}

\emph{Proof of proposition 1 is verbatim to the proof of statement 1 of proposition 5 from \cite{b16}.}

\textbf{Proposition 2.} \emph{If}
\begin{equation}\label{eq19}
\forall t \ge {t_0}{\rm{\;}}\int\limits_{{t_0}}^t {{\zeta _{ij}}\left( s \right){w_i}\left( s \right)ds}  < \infty {\rm{,\;}}\forall i{\rm{,\;}}j = 1{\rm{,}} \ldots {\rm{,}}{{{n}} _\theta }{\rm{,}}    
\end{equation}
\emph{then}
\begin{enumerate}
    \item[{\it i})] ${{\mathcal W}_i} \in {L_l}$ \emph{for any} $l \in \left( {1{\rm{,\;}}\infty } \right)$,
    \item[{\it ii})] \emph{there exists ${c_{\mathcal W}} > 0$ such that}
    \begin{equation}\label{eq20}
        \left| {{{\mathcal W}_i}\left( t \right)} \right| \le {c_{\mathcal W}}\frac{{\dot F\left( t \right)}}{{F\left( t \right)}} < \infty .
    \end{equation}
\end{enumerate}

\emph{Proposition 2 is a straightforward extension of statement 3 of proposition 2 from \cite{b16} to a multivariable case.}

Considering stationary signals ${\zeta _{ij}}\left( t \right)$ and ${w_i}\left( t \right)$, the boundedness of \eqref{eq19} is achieved in case the spectra of such signals have no common frequencies, which is not a restrictive condition for {\it truly} external perturbations that are independent with the reference trajectory ${q_d}\left( t \right)$. From the obtained linear regression equation \eqref{eq17}, the following choice of extra term for adaptation law \eqref{eq11} is proposed:
\begin{equation}\label{eq21}
{\vartheta _c}\left( t \right) = \gamma \Delta \left( t \right)\left( {{\mathcal Y}\left( t \right) - \Delta \left( t \right)\hat \theta \left( t \right)} \right){\rm{,}}  
\end{equation}
where $\gamma  > 0$ is a learning rate.

According to results of \cite{b16}, the following theorem holds for error $\tilde \theta \left( t \right)$.

\textbf{Theorem 1.} \emph{Assume that $r\left( t \right) = 0{\rm{,\;}}\forall t \ge {t_0}$ and:}
\begin{enumerate}
    \item[\textbf{C1)}] \emph{the inequality (19) holds,}
    \item[\textbf{C2)}] $\Delta  \notin {L_2}$.
\end{enumerate}
\emph{Then for all $i = 1{\rm{,}} \ldots {n_\theta }$ the error ${\tilde \theta _i}\left( t \right)$ converges to zero asymptotically.}

\emph{Proof of Theorem 1 statements can be found in \cite{b16}.}

In proposition 2 it is proved that, if PE-like inequality \eqref{eq18} holds, then the obtained scalar regressor $\Delta \left( t \right)$ is bounded away from zero. The inequality \eqref{eq18} means that both $\varphi \left( t \right)$ and $\zeta \left( t \right)$ are persistently exciting with sufficiently large excitation level. Therefore, premises of Theorem 1 to ensure asymptotic convergence are weaker in comparison with PE-like condition \eqref{eq18}, as, for example, there exists $\Delta \left( t \right){\rm{:}} = {\textstyle{1 \over {\sqrt {t + 1} }}}$, which is not square integrable and not PE as it vanishes to zero. As ${{\mathcal W}_i} \in {L_l}$ for any $l \in \left( {1{\rm{,\;}}\infty } \right)$ and $\Delta  \notin {L_2}$, then the interpretation of the convergence conditions from Theorem 1 is that the perturbation ${{\mathcal W}_i}\left( t \right)$ needs to converge to zero more rapidly in comparison with $\Delta \left( t \right)$.

\textbf{Remark 3.} In case when it is {\it a priori} known that the estimate of $\theta $ is consistent in the sense that $\left\| {\theta  - {{\hat \theta }_0}} \right\| \leqslant \epsilon$ for a sufficiently small $\epsilon > 0$, the choice 
\begin{equation*}
\begin{array}{l}
    \dot z = l\left( \left( {K + {\delta _\mu }\mu \left( t \right)} \right)r + {\Phi ^{\top}}\left( {q{\rm{,\;}}\dot q{\rm{,\;}}v{\rm{,\;}}\dot v} \right)\hat \theta  - z\right)
\end{array}  
\end{equation*}
ensures that the new perturbation in \eqref{eq13} satisfies
\begin{equation*}
\begin{array}{l}
\dot w = l\left(- w\underbrace { - {\tau _d}
-\mu \left( t \right)\left[ \Phi _M^{\top} - \Phi _{Mf}^{\top} -  {\textstyle{1 \over \mu\left(t\right) }}\Phi _f^{\top} \right]\hat \theta + {u_f}}_{ \approx f\left( t \right)- \hat f\left( t \right)} \right)    
\end{array}
\end{equation*}
and, therefore, it is small in the average sense in comparison with pure $ - {\tau _d}\left( t \right)$. However, such alternative choice could potentially violate the condition \eqref{eq19}.

\subsection{Stability analysis}

The conditions, under which the goal \eqref{eq2} is achieved when the control law \eqref{eq9} and estimation law \eqref{eq11}+\eqref{eq21} are used, are described in the following theorem.

\textbf{Theorem 2.} \emph{The robotic system \eqref{eq1} under properties 1-4 driven by the control law \eqref{eq9} with composite adaptation \eqref{eq11} + \eqref{eq21} is globally stable in the sense that:}
\begin{enumerate}
    \item[\textbf{1.}] \emph{If \textbf{C1} is met, then the tracking $e\left( t \right)$ and parametric $\tilde \theta \left( t \right)$ errors are bounded.}
    \item[\textbf{2.}] \emph{If \textbf{C1} and \textbf{C2} are satisfied, then both tracking and parametric errors converge to zero asymptotically.}
\end{enumerate}
\emph{Proof.} 

\textbf{1.} The following quadratic form is introduced:
\begin{equation}\label{eq23}
V = {\textstyle{1 \over 2}}{r^{\top}}Mr + {\textstyle{1 \over 2}}{\tilde \theta ^{\top}}{\Gamma ^{ - 1}}\tilde \theta  + {\textstyle{\rho  \over 2}}\tilde f_n^{\top}{\tilde f_n}.    
\end{equation}
According to \eqref{eq7}, the differential equation for ${\tilde f_n}\left( t \right)$ takes the form:
\begin{displaymath}\label{eq24}
\begin{array}{l}
{{\dot {\tilde f}}_n}= {\textstyle{1 \over {\mu \left( t \right)}}}\left( {\dot f - \dot \chi + \dot \Phi _f^{\top}\theta  + {{\dot u}_f}} \right)  
- {\textstyle{{\dot \mu \left( t \right)} \over {{\mu ^2}\left( t \right)}}}\tilde f =\\=  - {\textstyle{{\dot \mu \left( t \right)} \over {\mu \left( t \right)}}}{{\tilde f}_n} + {\textstyle{1 \over {\mu \left( t \right)}}}\dot f - \dot x + {{\dot x}_f} - {\textstyle{{\dot \mu \left( t \right)} \over {\mu \left( t \right)}}}\left( {x - {x_f}} \right) + u - {u_f} + \\
 +\! \left( {{\Phi ^{\top}}\!\left( {q{\rm{,\;\!}}\dot q{\rm{,\;\!}}v{\rm{,\;\!}}\dot v} \right) \!-\! \Phi _C^{\top}\!\left( {q{\rm{,\;\!}}\dot q{\rm{,\;\!}}v} \right) \!+\! \Phi _{\partial M}^{\top}\!\left( {q{\rm{,\;\!}}\dot q{\rm{,\;\!}}r} \right) \!-\! {\Phi _f}} \right)\!\theta  \!= \\
 =  - {\textstyle{{\dot \mu \left( t \right)} \over {\mu}\left( t \right)}}{{\tilde f}_n} + {\textstyle{1 \over \mu\left( t \right) }}\dot f -\\
 -\! \left( {{\Phi ^{\top}}\!\left( {q{\rm{,\;}}\dot q{\rm{,\;}}v{\rm{,\;}}\dot v} \right) - \Phi _C^{\top}\left( {q{\rm{,\;}}\dot q{\rm{,\;}}r} \right) + \Phi _{\partial M}^{\top}\left( {q{\rm{,\;}}\dot q{\rm{,\;}}r} \right)} \right)\theta  - \\
 - f - u + \chi + u - {u_f}+ \\
 +\! \left( {{\Phi ^{\top}}\!\left( {q{\rm{,\;\!}}\dot q{\rm{,\;\!}}v{\rm{,\;\!}}\dot v} \right) \!-\! \Phi _C^{\top}\!\left( {q{\rm{,\;\!}}\dot q{\rm{,\;\!}}v} \right) \!+\! \Phi _{\partial M}^{\top}\!\left( {q{\rm{,\;\!}}\dot q{\rm{,\;\!}}r} \right) \!-\! {\Phi _f}} \right)\!\theta \! = \\
 =  - {\textstyle{{\dot \mu \left( t \right)} \over {\mu \left( t \right)}}}{{\tilde f}_n} + {\textstyle{1 \over {\mu \left( t \right)}}}\dot f - f
 + \chi - {\Phi _f}\theta  - {u_f} = \\
 =  - {\textstyle{{{\mu ^2}\left( t \right) + \dot \mu \left( t \right)} \over {\mu \left( t \right)}}}{{\tilde f}_n} + {\textstyle{1 \over {\mu \left( t \right)}}}\dot f{\rm{,\;}}{{\tilde f}_n}\left( {{t_0}} \right) = {0_n}.
\end{array}    
\end{displaymath}
Then the derivative of \eqref{eq23} is written as:
\begin{equation}\label{eq25}
\begin{array}{l}
\dot V \!=\! {r^{\top}}\!\left( \!{ - \left( {K + {\delta _\mu }\mu } \right)r \!-\! C\left( {q{\rm{,\;}}\dot q} \right)r \!+\! {\Pi ^{\top}}\tilde \theta  \!+\! \mu {{\tilde f}_n}}\! \right) \!+ \\
+ {\textstyle{1 \over 2}}{r^{\top}}{\textstyle{d \over {dt}}}\left[ {M\left( q \right)} \right]r +  \tilde \theta {\Gamma ^{ - 1}}\dot {\tilde \theta}  + \rho \tilde f_n^{\top}{{\dot {\tilde f}}_n} = \\
 = \! -  {r^{\top}}\!\left(\! {K \!+\! {\delta _\mu }\mu } \right)r \!+\! {r^{\top}}\!\left[ {{\textstyle{1 \over 2}}\dot M \!-\! C\left( {q{\rm{,\;}}\dot q} \right)} \right]\!r \!+\! \mu {r^{\top}}\!{{\tilde f}_n}\! +\! \\
 + {{\tilde \theta }^{\top}}\!\left( { - \gamma {\Delta ^2}\tilde \theta  - \gamma \Delta {\mathcal W}} \right) \!-\! \rho {\textstyle{{{\mu ^2} + \dot \mu } \over \mu }}\tilde f_n^{\top}{{\tilde f}_n} + {\textstyle{\rho  \over \mu }}\tilde f_n^{\top}\dot f \!=\! \\
 \!=\! - {r^{\top}}\left( {K + {\delta _\mu }\mu } \right)r - \gamma {\Delta ^2}{{\tilde \theta }^{\top}}\tilde \theta  - \rho {\textstyle{{{\mu ^2} + \dot \mu } \over \mu }}\tilde f_n^{\top}{{\tilde f}_n} + \\
 + \mu {r^{\top}}{{\tilde f}_n} - \gamma {{\tilde \theta }^{\top}}\Delta {\mathcal W} + {\textstyle{\rho  \over \mu }}\tilde f_n^{\top}\dot f.
\end{array}    
\end{equation}
To further prove the system stability, the following upper bounds are obtained with the help of the inequality \linebreak $ab \le 2ab \le \delta {a^2} + {\delta ^{ - 1}}{b^2}$:
\begin{equation*}
\begin{array}{l}
\mu \left\| r \right\|\left\| {{{\tilde f}_n}} \right\| \le {\delta _\mu }\mu {\left\| r \right\|^2} + \delta _\mu ^{ - 1}\mu {\left\| {{{\tilde f}_n}} \right\|^2}{\rm{,}}\\
{\textstyle{\rho  \over \mu }}\left\| {{{\tilde f}_n}} \right\|\left\| {\dot f} \right\| \!=\! {\textstyle{\rho  \over \mu }}\left\| {{{\tilde f}_n}} \right\|\left\| {{{\dot \tau }_d}} \right\| \le \rho {\delta _\rho }{\left\| {{{\dot \tau }_d}} \right\|^2}{\mu ^{ - 2}}
+ \rho \delta _\rho ^{ - 1}{\left\| {{{\tilde f}_n}} \right\|^2}\!{\rm{,}}\\
\gamma \left\| {\tilde \theta } \right\|\left| \Delta  \right|\left\| {\mathcal W} \right\| \le \gamma {\delta _{\mathcal W}}{\Delta ^2}{\left\| {\tilde \theta } \right\|^2} + \gamma \delta _{\mathcal W}^{ - 1}{\left\| {\mathcal W} \right\|^2}{\rm{,}}
\end{array}    
\end{equation*}
where ${\delta _{\mathcal W}} \in \left( {0{\rm{, 1}}} \right)$.

Then, according to property 4, the right-hand side of derivative \eqref{eq25} could be bounded from above:
\begin{equation}\label{eq27}
\begin{array}{l}
\dot V \le  - {\lambda _{{\rm{min}}}}\left( K \right){\left\| r \right\|^2} - \gamma {\Delta ^2}\left[ {1 - {\delta _{\mathcal W}}} \right]{\left\| {\tilde \theta } \right\|^2} - \\
 - \left[ {\left( {\rho  - \delta _\mu ^{ - 1}} \right)\mu  + \rho {\textstyle{{\dot \mu } \over \mu }} - \rho \delta _\rho ^{ - 1}} \right]{\left\| {{{\tilde f}_n}} \right\|^2} + \\
 + \gamma \delta _{\mathcal W}^{ - 1}{\left\| {\mathcal W} \right\|^2} + \rho {\delta _\rho }{\mu ^{ - 2}}{\left\| {{{\dot \tau }_d}} \right\|^2} \le \\
 \le  - {\lambda _{{\rm{min}}}}\left( K \right){\left\| r \right\|^2} - \gamma {\Delta ^2}\left[ {1 - {\delta _{\mathcal W}}} \right]{\left\| {\tilde \theta } \right\|^2} - \\
 - {\mu _{LB}}{\left\| {{{\tilde f}_n}} \right\|^2} + \gamma \delta _{\mathcal W}^{ - 1}{\left\| {\mathcal W} \right\|^2} + \rho {\delta _\rho }{\mu ^{ - 2}}{\left\| {{{\dot \tau }_d}} \right\|^2}.
\end{array}    
\end{equation}
As ${\lambda _{{\rm{min}}}}\left( K \right) > 0$ and ${\mu _{{\rm{LB}}}}\left( t \right) > 0{\rm{,\;}}{\delta _{\mathcal W}} \in \left( {0{\rm{,\;1}}} \right)$, then the derivative \eqref{eq27} is rewritten as follows:
\begin{equation}\label{eq28}
\dot V\left( t \right) \!\le\!  - {\eta ^2}\left( t \right)V\left( t \right) + \gamma \delta _{\mathcal W}^{ - 1}{\left\| {{\mathcal W}\left( t \right)} \right\|^2} + \rho {\delta _\rho }{\left\| {\lambda \left( t \right)} \right\|^2}{\rm{,}}    
\end{equation}
where ${\eta ^2}\!\left( t \right)\!{\rm{:}} \!=\! {\rm{min}}\left\{ {{\textstyle{{{\lambda _{{\rm{min}}}}\left( K \right)} \over {\overline m}}}{\rm{, }}{\textstyle{{2\gamma {\Delta ^2}\left( t \right)\left( {1 - {\delta _{\mathcal W}}} \right)} \over {{\lambda _{{\rm{max}}}}\left( {{\Gamma ^{ - 1}}} \right)}}}{\rm{, 2}}{\textstyle{{{\mu _{{\rm{LB}}}}\left( t \right)} \over \rho }}} \right\} \!\ge \!0.$

In the worst case $\eta \left( t \right) = 0$, according to comparison lemma, for all $t \ge {t_0}$ the solution of \eqref{eq28} is obtained as:
\begin{equation*}
\begin{array}{l}
V\left( t \right) \le V\left( {{t_0}} \right) + \gamma \delta _{\mathcal W}^{ - 1}\int\limits_{{t_0}}^t {{{\left\| {{\mathcal W}\left( s \right)} \right\|}^2}} ds 
+ \rho {\delta _\rho }\int\limits_{{t_0}}^t {{{\left\| {\lambda \left( s \right)} \right\|}^2}} ds \\ =\! V\left( {{t_0}} \right)
+ \gamma \delta _{\mathcal W}^{ - 1}\int\limits_{{t_0}}^t {\sum\limits_{i = 1}^{{n_\theta }} {{\mathcal W}_i^2\left( s \right)} } ds + \rho {\delta _\rho }\int\limits_{{t_0}}^t {{{\left\| {\lambda \left( s \right)} \right\|}^2}ds}  \!<\! \infty.
\end{array}    
\end{equation*}
As condition \textbf{C1)} holds, then ${{\mathcal W}_i} \in {L_2}$ according to proposition 2, and therefore, application of property 4 shows that the function $V\left( t \right)$ defined as \eqref{eq23} is bounded from above. Therefore, using standard arguments \cite{b1}, the tracking $e\left( t \right)$ and parametric $\tilde \theta \left( t \right)$ errors are also bounded, which completes proof of the first part of Theorem 2.

\textbf{2.} To prove the second statement, equation \eqref{eq28} is rewritten in the following form:
\begin{equation}\label{eq30}
\dot V\left( t \right) \leqslant  - {\eta ^2}\left( t \right)V\left( t \right) + \epsilon\left( t \right){\text{,}}  
\end{equation}
where $\epsilon\left( t \right) = \gamma \delta _\mathcal{W}^{ - 1}\sum\limits_{i = 1}^{{n_\theta }} {\mathcal{W}_i^2\left( t \right)}  + \rho {\delta _\rho }{\left\| {\lambda \left( t \right)} \right\|^2}.$

For further analysis, we recall the following useful Lemma \cite{b17}.

\textbf{Lemma 1.} \emph{Consider the scalar system defined by}
\begin{displaymath}
    \dot x\left( t \right) =  - {a^2}\left( t \right)x\left( t \right) + b\left( t \right){\rm{,\;}}x\left( {{t_0}} \right) = {x_0}{\rm{,}}
\end{displaymath}
\emph{where $x\left( t \right) \in \mathbb{R}{\rm{,\;}}a{\rm{,\;}}b{\rm{:\;}}{\mathbb{R}_+ } \mapsto \mathbb{R}$ are piecewise continuous bounded functions. If $a \notin {L_2}$ and $b \in {L_1}$, then }
\begin{equation*}
    \mathop {{\rm{lim}}}\limits_{t \to \infty } x\left( t \right) = 0.
\end{equation*}
As property 4 and conditions \textbf{C1)}-\textbf{C2)} hold, then $\eta  \notin {L_2}$ and $\epsilon\in {L_1}$. Therefore, according to Comparison Lemma and Lemma 1, the following limit exists:
\begin{equation}\label{eq31}
\mathop {{\rm{lim}}}\limits_{t \to \infty } V\left( t \right) = 0{\rm{,}}    
\end{equation}
which implies that both filtered tracking $r\left( t \right)$ and parametric $\tilde \theta \left( t \right)$ errors asymptotically converge to zero. 

To prove that the error $e\left( t \right)$ also converges to zero, inequality \eqref{eq30} is integrated to obtain:
\begin{displaymath}
V\left( t \right) \leqslant V\left( {{t_0}} \right) - \int\limits_{{t_0}}^t {{\eta ^2}\left( s \right)V\left( s \right)ds}  + \int\limits_{{t_0}}^t {\epsilon\left( s \right)ds} .
\end{displaymath}
As $V \in {L_\infty }$ (from proof of first part of theorem) and $\epsilon \in {L_1}$, then the following signal is bounded:
\begin{displaymath}
\int\limits_{{t_0}}^t {{\eta ^2}\left( s \right)V\left( s \right)ds}  \leqslant V\left( {{t_0}} \right) - V\left( t \right) + \int\limits_{{t_0}}^t {\epsilon\left( s \right)ds}  < \infty {\text{,}}
\end{displaymath}	
from which we have (here it is assumed that ${\eta ^2}\left( t \right) > 0$ almost everywhere and, therefore, ${\rm{ess}}\mathop {{\rm{inf}}}\limits_t {\eta ^2}\left( t \right) \ne 0$):
\begin{displaymath}
{\rm{ess}}\mathop {{\rm{inf}}}\limits_t {\eta ^2}\left( t \right)\int\limits_{{t_0}}^t {V\left( s \right)ds}  \le \int\limits_{{t_0}}^t {{\eta ^2}\left( s \right)V\left( s \right)ds}  < \infty {\rm{,}}
\end{displaymath}
which means $r \in {L_2}$, and, therefore, applying Lemma 2.18 from \cite{b5}, we have $\mathop {{\rm{lim}}}\limits_{t \to \infty } e\left( t \right) = {\rm{0}}{\rm{.}}$

\emph{Proof of Theorem 2 is completed.}

\textbf{Remark 4.} Generally, to achieve the goal of asymptotic tracking \eqref{eq2}, the proposed composite adaptive disturbance rejection control scheme needs to satisfy three key assumptions: \textbf{KA1)} $\Delta  \notin {L_2}$, \textbf{KA2)} $\lambda  \in {L_2}$ and \linebreak \textbf{KA3)} the external torque ${\tau _d}\left( t \right)$ and instrumental variable $\zeta \left( t \right)$ are independent in the sense of inequality \eqref{eq19}. Assumption $\Delta  \notin {L_2}$ is weaker in comparison with PE-like condition \eqref{eq18}, and the requirements \textbf{KA2)}, \textbf{KA3)} refer to robot environment and high-level tracking objectives. The proposed control system \eqref{eq9} + \eqref{eq11} + \eqref{eq21} ensures asymptotic stability \eqref{eq2}, which is more preferable property in comparison with UUB/ISS with arbitrarily bad bounds, which are guaranteed by the existing approaches \cite{b10,b11,b12,b13,b14,b15}. However, unfortunately, the condition $\Delta  \notin {L_2}$ for convergence is stronger in comparison with initial/finite excitation requirement, which is thoroughly exploited in the studies \cite{b10,b11,b12,b13,b14,b15}.

\section{Numerical Experiments}\label{s6}

The following model of a two link manipulator has been considered:
\begin{equation*}
\begin{array}{c}
M\left( q \right) =  {\begin{bmatrix}
{{\theta _1} + 2{\theta _2}{{\rm{c}}_2}}&{{\theta _3} + {\theta _2}{{\rm{c}}_2}}\\
{{\theta _3} + {\theta _2}{{\rm{c}}_2}}&{{\theta _3}}
\end{bmatrix}} {\rm{, }}\\
C\left( {q{\rm{,\;}}\dot q} \right) = {\begin{bmatrix}
{ - {\theta _2}{{\rm{s}}_2}{{\dot q}_2}}&{ - {\theta _2}{{\rm{s}}_2}\left( {{{\dot q}_1} + {{\dot q}_2}} \right)}\\
{{\theta _2}{{\rm{s}}_2}{{\dot q}_1}}&0
\end{bmatrix}}{\rm{, }}\\
F\left( {\dot q} \right) + G\left( q \right) = {\begin{bmatrix}
{{\theta _4}g{{\rm{c}}_{12}} + {\theta _5}g{{\rm{c}}_1}}\\
{{\theta _4}g{{\rm{c}}_{12}}}
\end{bmatrix}}{\rm{, }}
\end{array}    
\end{equation*}
where ${{\rm{c}}_1} = {\rm{cos}}\left( {{q_1}} \right){\rm{,}}\;{{\rm{c}}_2} = {\rm{cos}}\left( {{q_2}} \right){\rm{,\;}}{{\rm{s}}_2} = {\rm{sin}}\left( {{q_2}} \right){\rm{,\;}}\linebreak {{\rm{c}}_{12}} = {\rm{cos}}\left( {{q_1} + {q_2}} \right)$, and the parameters of the system were picked as follows for the numerical experiments:
\begin{displaymath}
{\theta _1} = 1.3{\rm{,\;}}{\theta _2} = 0.28{\rm{,\;}}{\theta _3} = 0.32{\rm{,\;}}{\theta _4} = 0.4{\rm{,\;}}{\theta _5} = 1.4. 
\end{displaymath}
The regressors from property 1 were defined in the following way:
\begin{displaymath}\label{eq33}
\begin{array}{l}
\Phi _M^{\top}\left( {q{\rm{,\;}}\dot v} \right) = {\begin{bmatrix}
{\begin{array}{*{20}{c}}
{{{\dot v}_1}}&{{{\rm{c}}_2}\left( {2{{\dot v}_1} + {{\dot v}_2}} \right)}&{{{\dot v}_2}}\\
0&{{{\rm{c}}_2}{{\dot v}_1}}&{{{\dot v}_1} + {{\dot v}_2}}
\end{array}}&{{0_{2 \times 2}}}
\end{bmatrix}}{\rm{, }}\\
\Phi _C^{\top}\left( {q{\rm{,\;}}\dot q{\rm{,\;}}v} \right) = {\begin{bmatrix}
{{0_{2 \times 1}}}&{\begin{array}{*{20}{c}}
{ - {{\rm{s}}_2}\left( {{{\dot q}_2}{v_1} + \left( {{{\dot q}_1} + {{\dot q}_2}} \right){v_2}} \right)}\\
{{{\rm{s}}_2}{{\dot q}_1}{v_1}}
\end{array}}&{{0_{2 \times 3}}}
\end{bmatrix}}{\rm{, }}\\
\Phi _F^{\top}\left( {\dot q} \right) + \Phi _G^{\top}\left( q \right) = {\begin{bmatrix}
{{0_{2 \times 3}}}&{\begin{array}{*{20}{c}}
{g{{\rm{c}}_{12}}}&{g{{\rm{c}}_1}}\\
{g{{\rm{c}}_{12}}}&0
\end{array}}
\end{bmatrix}}{\rm{,}}\\
\Phi _{\partial M}^{\top}\left( {q{\rm{,\;}}\dot q{\rm{,\;}}\dot v} \right) = {\begin{bmatrix}
{{0_{2 \times 1}}}&{\begin{array}{*{20}{c}}
{ - {{\rm{s}}_2}{{\dot q}_2}\left( {2{{\dot v}_1} + {{\dot v}_2}} \right)}\\
{ - {{\rm{s}}_2}{{\dot q}_2}{{\dot v}_1}}
\end{array}}&{{0_{2 \times 3}}}
\end{bmatrix}}.
\end{array}    
\end{displaymath}
The filtered version $\varphi \left( t \right)$ of regressor ${\Phi ^{\top}}\left( {q{\rm{,\;}}\dot q{\rm{,\;}}\ddot q} \right)$ was computed without acceleration $\ddot q$ according to \cite[(6.6.9)]{b1} and satisfied the relationship:
\begin{equation*}\label{eq34}
\begin{gathered}
    \varphi \!\left( t \right)\!{\rm{:}} \!=\! {\textstyle{s \over {ls + 1}}}\left[ {\Phi _M^{\top}}\right]\!\left(t\right) \!+\! {\textstyle{1 \over {ls + 1}}}\!\left[  - \Phi _{\partial M}^{\top} \!+\! \Phi _C^{\top} \!+\! \Phi _F^{\top} + \Phi _G^{\top} \right]\!\left(t\right).
\end{gathered}
\end{equation*}
For comparison purposes, we also implemented the following composite adaptation law \cite{b12}:
\begin{equation}\label{eq35}
\begin{array}{l}
\dot {\hat \theta} \left( t \right) = \Gamma \left( {{\Pi ^{\top}}\left( t \right)r\left( t \right) + {{\hat \theta }_c}\left( t \right)} \right){\rm{,\;}}\hat \theta \left( {{t_0}} \right) = {{\hat \theta }_0}{\rm{,}}\\
{{\hat \theta }_c}\left( t \right) = \gamma \left( {Y\left( {{t_e}} \right) - \Psi \left( {{t_e}} \right)\hat \theta \left( t \right)} \right){\rm{,}}
\end{array}    
\end{equation}
where
\begin{displaymath}
\begin{array}{l}
\dot Y\left( t \right) = \varphi \left( t \right)z\left( t \right) - \varphi \left( {t - T} \right)z\left( {t - T} \right){\rm{, }}\\
\dot \Psi \left( t \right) = \varphi \left( t \right){\varphi ^{\top}}\left( t \right) - \varphi \left( {t - T} \right){\varphi ^{\top}}\left( {t - T} \right){\rm{, }}\\
Y\left( {{t_0}} \right) = {0_{{n_\theta }}}{\rm{,\;}}\psi \left( {{t_0}} \right) = {0_{{n_\theta } \times {n_\theta }}}{\rm{,}}\\
\dot z\left( t \right) = l\left( {\tau \left( t \right) - z\left( t \right)} \right){\rm{,\;}}z\left( {{t_0}} \right) = {0_n}{\rm{,}}\\
\dot \varphi \left( t \right) = l\left( {\Phi \left( {q{\rm{,\;}}\dot q{\rm{,\;}}\ddot q} \right) - \varphi \left( t \right)} \right){\rm{,\;}}\varphi \left( {{t_0}} \right) = {0_{{n_\theta } \times n}}
\end{array}    
\end{displaymath}
and
\begin{displaymath}
{t_e}{\rm{:}} = {\rm{arg }}\mathop {{\rm{max}}}\limits_{\forall t} {\lambda _{{\rm{min}}}}\left( {\Psi \left( t \right)} \right).   
\end{displaymath}
The parameters of the control law \eqref{eq9}, filters \eqref{eq7}, \eqref{eq13}, \eqref{eq15a}, \eqref{eq15b} and estimation laws \eqref{eq11} + \eqref{eq21}, \eqref{eq35} were chosen as follows:
\begin{equation*}\label{eq36}
\begin{array}{c}
\alpha  = 1{\rm{,\;}}K = 2{I_2}{\rm{,\;}}l = 50{\rm{,\;}}p = 2{\rm{,\;}}T = 20{\rm{,}}\;{\delta _\mu } = 0.8{\rm{,}}\\
\Gamma  = 0.01{I_5}{\rm{,\;}}\gamma  = \left\{ \begin{array}{l}
{\rm{100}} \cdot {10^{8}}{\rm{,\;for\;}} \eqref{eq21} \\
{\rm{100}} \cdot {10^{ - 2}}{\rm{,\;for\;}}\eqref{eq35}
\end{array} \right..
\end{array}    
\end{equation*}
The value of the gain for estimation law \eqref{eq11} + \eqref{eq21} was chosen so large due to the small order of the signal ${\Delta ^2}\left( t \right)$ amplitude.
The desired trajectory and initial conditions for the system were chosen as follows:
\begin{displaymath}
\begin{array}{l}
{q_{1d}}\left( t \right) = 0.4\pi {\rm{sin}}\left( {2t} \right) + 0.2\pi {\rm{,\;}} {q_{1}}\left( t_{0} \right) = 0{\rm{,}} \\
{q_{2d}}\left( t \right) = 0.3\pi {\rm{sin}}\left( {0.3t + {\textstyle{\pi  \over 2}}} \right) + 0.3\pi{\rm{,\;}} {q_{2}}\left( t_{0} \right) = 0.3\pi.
\end{array}    
\end{displaymath}
The external torque and weighted function were chosen as follows
\begin{displaymath}
\begin{array}{*{20}{c}}
{{\tau _{1d}}\left( t \right) = 7.5{\rm{sin}}\left( {0.5\pi t} \right)}\\
{{\tau _{2d}}\left( t \right) = 1.5{\rm{sin}}\left( {0.05\pi t} \right)}
\end{array}{\rm{,\;}}\mu\left(t\right)  = t + 15{\rm{, }}    
\end{displaymath}
which ensured that the property 4 was met.

Figure 1 depicts the behavior of the regressor $\Delta \left( t \right)$ and disturbance ${\mathcal W}\left( t \right).$
\begin{figure}[h]
      \centering
      \includegraphics[scale=0.6]{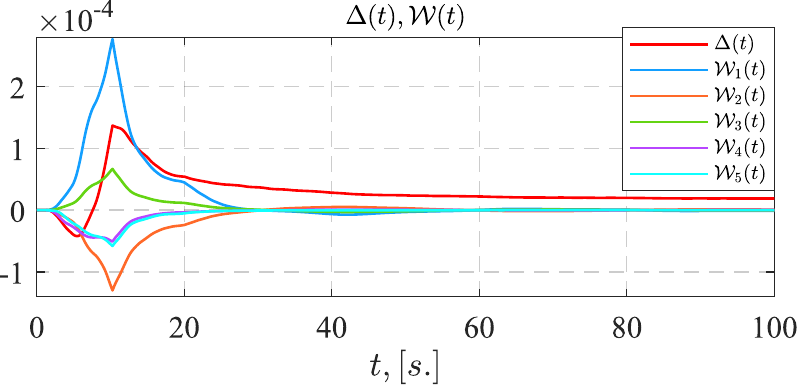}
      \caption{Regressor $\Delta \left( t \right)$ and disturbance ${\mathcal W}\left( t \right)$ behavior.}
      \label{Figure1} 
      \end{figure}

As the regressor $\Delta \left( t \right)$ was vanishing slower than the disturbance ${\mathcal W}\left( t \right)$, then it was concluded that the condition of the asymptotic stability from statement 2 of theorem 2 was met.

\begin{figure}[h]
      \centering
      \includegraphics[scale=0.58]{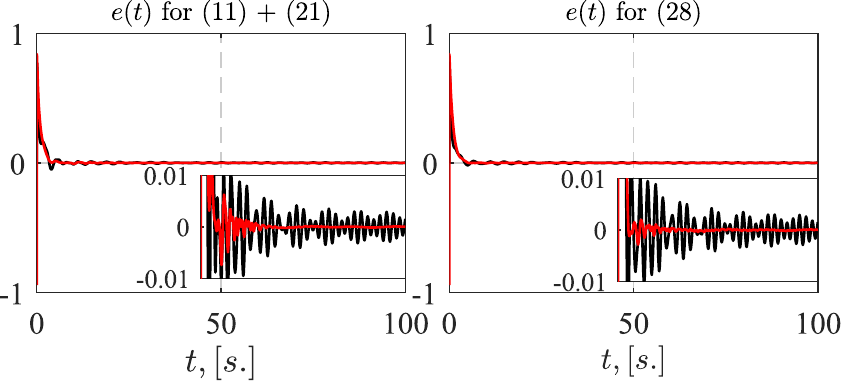}
      \caption{Behavior of $e\left( t \right)$ for \eqref{eq11} + \eqref{eq21} and \eqref{eq35}.}
      \label{Figure2} 
      \end{figure}

\begin{figure}[h]
      \centering
      \includegraphics[scale=0.58]{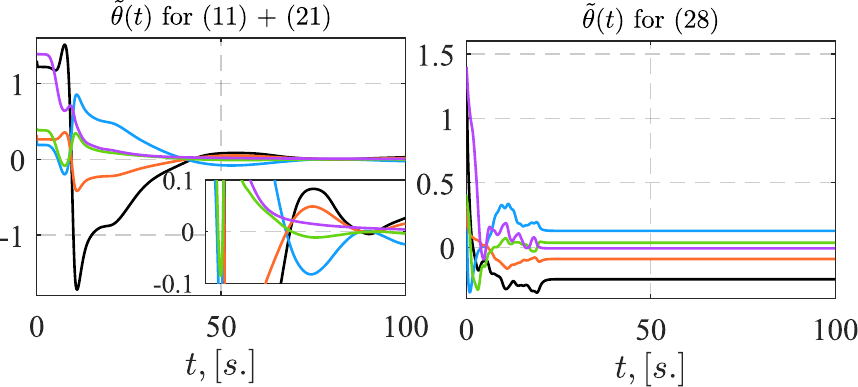}
      \caption{Behavior of $\tilde \theta \left( t \right)$ for \eqref{eq11} + \eqref{eq21} and \eqref{eq35}.}
      \label{Figure3} 
      \end{figure}

\begin{figure}[h]
      \centering
      \includegraphics[scale=0.55]{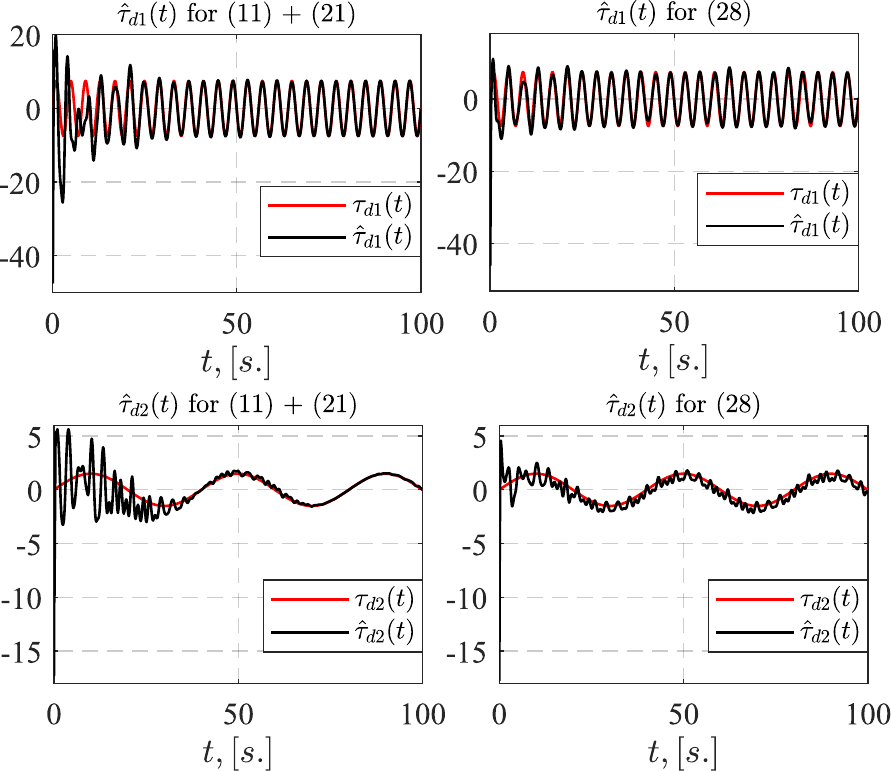}
      \caption{Behavior of ${\tilde \tau _d}\left( t \right)$ for \eqref{eq11} + \eqref{eq21} and \eqref{eq35}.}
      \label{Figure4} 
      \end{figure}

Figures \ref{Figure2}-\ref{Figure4} presents the comparison of parametric errors $\tilde \theta \left( t \right)$, tracking errors $e\left( t \right)$ and external disturbance reconstruction errors ${\tilde \tau _d}\left( t \right) = {\tau _d}\left( t \right) - {\hat \tau _d}\left( t \right)$ for the algorithms \eqref{eq11} + \eqref{eq21} and \eqref{eq35}. Here 
\begin{equation*}
\begin{array}{l}
{\hat \tau _d}\left( t \right) \!=\! -\mu\left(t\right) \left[ {\Phi _M^{\top}\left( {q{\rm{,\;}}r} \right) \!-\! \Phi _{Mf}^{\top}\left( t \right) \!-\! {\textstyle{1 \over \mu\left(t\right) }}\Phi _f^{\top}\left( t \right)} \right]\hat \theta \left( t \right) \!+ \\ \hfill +{u_f}\left( t \right) 
\end{array}
\end{equation*}
denotes the adaptive estimation of the external disturbance extracted from the control signal \eqref{eq9}.

The results presented in Fig. \ref{Figure2}-\ref{Figure4} allow one to make the following conclusions:
\begin{enumerate}
    \item[\textbf{C1)}] the adaptation law \eqref{eq35} guarantees boundedness of the unknown parameter estimation error in the presence of an external perturbation,
    \item[\textbf{C2)}] the adaptation law \eqref{eq11} + \eqref{eq21} ensures asymptotic convergence to zero of the unknown parameter estimation error in presence of external perturbation,
    \item[\textbf{C3)}] Disturbance Rejection control law \eqref{eq9} augmented with the proposed adaptation law \eqref{eq11} + \eqref{eq21} provides better estimation accuracy of the external torque in comparison with the Disturbance Rejection control law \eqref{eq9} augmented with the existing solution \eqref{eq35}.
\end{enumerate}


\section{Conclusion}\label{s7}
In this paper, a novel composite adaptation scheme is proposed for robotics systems affected by unknown external perturbations, which guarantees achievement of the global stability objective regardless of the parametric error convergence and, additionally, asymptotic convergence of the parameter estimation and tracking errors to zero in case the following non-restrictive (in authors’ opinion) assumptions hold: ({\it i}) the weighted rate of the external torque is square integrable, ({\it ii}) the external torque affecting the plant is independent with the system trajectory, ({\it iii}) the scalar regressor is not integrable with square. Results of numerical simulations corroborated the theoretical results and demonstrated advantages of the proposed approach.

\bibliographystyle{plain}        
\bibliography{autosam}           



\end{document}